\documentclass[aps,prl,twocolumn,showpacs]{revtex4}
\usepackage{graphics}
\usepackage{bm}
\usepackage{amsfonts}
\usepackage{amssymb}

\begin{document}
\newcommand{\eq}{\end{equation} }
\draft
%

\title{Dynamics of Scale Free Random Threshold Network}
\author{Ikuo Nakamura}
 \email{ikuo@arch.sony.co.jp}
\affiliation{Sony Corporation, 2-10-14 Osaki, Shinagawa, Tokyo, Japan}
\affiliation{Computer Science Department, Stanford University, CA 94305, 
USA}

\begin{abstract}
We study the dynamics of Random Threshold Network (RTN) on scale free networks, 
with asymmetric links, some interaction rules where propagation of local 
perturbations depends on in-degree $k$ of the nodes.
We find that there is no phase transition with respect to average connectivty 
independently of network topology for the case 
temperature $T=0$, threshold $h=0$ and 
the probability distribution of indegree $P(k)$ satisfies $P(0)=D=0$.
We have investigated the emergence of phase transition involving three parameters,
 i.e. $T,h$ and $D$. RTN can be continuously connected to 
Random Boolean Network (RBN) in $T\rightarrow \infty$,
and we find moderate thermal noise extends the regime of ordered dynamics,
 compared to RTN in $T=0$ regime and RBN. Furthermore, we discuss the dynamic 
properties from another point of view, dynamical mean field reaction rate equation.
\end{abstract}

\maketitle
	
Recently there has been attracted considerable interest 
in the formation of random networks, in the connectivity and  
especially in the dynamics of these networks \cite{WaSt,AlBa,BaAlRev}.
Many naturally occuring networks such as the Internet \cite{AlBar},
 WWW \cite{CaMa}, gene regulatory network\cite{Ag}, protein networks \cite{Je,MaSn} 
exhibit a power-law; i.e. scale free degree
distribution for the links. 
A scale free network have a distinctive property that a small
fraction of the nodes which is often called 'hub' are highly connected
whereas the majority of the nodes have low connectivities.
The absence of a typical scale of the
connectivity is often related to the organization of the network as a
hierarchy.
Although their topological properties have been studied in detail,
dynamical behavior provided with asymmetric links and some interaction 
rules are not sufficiently explored.
Especially in neural networks and gene regulatory networks,
interaction rules between nodes plays an important role to organize
the dynamical properties of the network.
Such a dynamical network model, e.g. Random Boolean 
Network (RBN) is first introduced by Kauffman \cite{Ka,Ka2} to model the gene
regulatory netoworks in the biological system.
In this models the state $\sigma_i\in \{0,1\}$ of a
network node $i$ is a logical function of the states of in-degree $k_i$ other nodes chosen at random.
The logical functions are chosen at random, with a suitably biased probability.
RBN model shows a phase transition with respect to the number
of inputs per node $K$ at a critical average connectivity $\langle K_c \rangle=[2p(1-p)]^{-1}$ from ordered to 
chaotic state, where $p$ is the probability that the randomly chosen output of node $i$ 
are unity.
In chaotic regime ($K>K_c$), a small perturbation in the initial state 
propagates across the entire system, whereas all perturbation in the initial 
state dies out in ordered regime ($K<K_c$).
The dynamics of RBN on scale free is recently investigated \cite{Aldana}. 
However, its dynamical property is the same as that of classical 
RBN model because the output state of the nodes in RBN is 
independent of their in-degree distribution. 
In order to study the dynamical effects of the in-degree (scale free) distribution,
we focus on Random Threshold Network (RTN),
first investigated as diluted and asymmetric spinglass models \cite{Derrida}, and asymmetric
neural networks \cite{DeGa,KrZi}. 
It is a subset of Boolean Network which is thought to show dynamical behavior similar to RBN 
\cite{BoRo}.

We consider a network of $N$ randomly interconnected binary nodes with state $\sigma_i=\pm1$. 
In this model, probability distribution of outputs strongly 
depends on their in-degree parameter.
At time $t$, the fields $f_i(t)$ of node $i$ are computed:
\begin{eqnarray}
f_i(t)&=&\sum_{j=1}^N c_{ij}\sigma_j(t)+h
\end{eqnarray}
where $h(\geq 0)$ is the threshold parameter.
Threshold characteristics can be often shown in neural networks and gene regulatory networks, i.e.
threshold for neuron firing and that for gene expression.
In this brief report, The interaction weight $c_{ij}$ take discrete values
$c_{ij}=\pm1$, with equal posiibility.
If node $i$ does not take a signal from $j$, one has $c_{ij}=0$.
We assume $c_{ii}$ is always zero to avoid $self$ $connection$.
The in-degree $k_i$ of node $i$ is defined as $k_i=\sum_j^N 
|c_{ij}|$. The interaction weights $c_{ij}$ is chosen randomly from all nodes 
but follows the in-degree distribution
$P(k)$. 
For each node $i$, its state at time $t+1$ are functions of the inputs it 
receives from other nodes at time $t$.
\begin{eqnarray}
\sigma_i(t+1)&=&\left\{
 \begin{array}{rl}
 1,& \quad \mbox{with probability $G(f_i(t))$}\\
 -1,& \quad \mbox{with probability $G(-f_i(t))$}
 \end{array}\right.
\end{eqnarray}
where $G(x)=[1+\mbox{exp}(-2x/T)]^{-1}$.
The connection between RBN and RTN is continuously made by introducing
 thermal noise to probability distribution of outputs function $G(x)$.
The parameter $T$ defines the temperature of the system.
The $N$ network nodes are idealized by synchronized updating.
Generalizing so-called Annealing Approximation method introduced by
Derrida and Pomeau \cite{DePo}, 
we approach the system analytically in the limit of a large system size $N$.
One finds that normalized overlap function $x$ obey the equation
\begin{eqnarray}
x(t+1)&=&F(x(t))
\label{overlapeq}
\end{eqnarray}
where the mapping function $F(x)$ is given by
\begin{eqnarray}
F(x)&\equiv&1-\sum_k^{\tilde{K}}p_s(k,h,T)P(k)(1-x^k)
\label{overlap}
\end{eqnarray}
Normalized overlap function 
$x(t)=1-d(\vec{\sigma(t)},\vec{\sigma'(t)})/N$ 
is a subtraction of the normalized Hamming distance 
 from unity where Hamming distance $d(\vec{\sigma(t)},\vec{\sigma'(t)})$ denotes overlap between time independent
 two distinctive configurations. 
In the $N\rightarrow \infty$ limit, $x$ represent the probability of two arbitary configurations
to be equal.
The possibility $p_s$ is output state reversal 
by changing the state of a single input $j$.
Using combinatorial methods \cite{RoBo}, the stochastic 
distribution of $p_s$ is derived.
\begin{widetext}
\begin{eqnarray}
p_s(k,h,T)=&&\frac{1}{k\cdot 2^k}\sum_{i=0}^k {k \choose i}
\biggl( G(k+h-2i)\bigl(k-(k-i)G(k+h-2-2i)-iG(k+h+2-2i)\bigr)\nonumber \\
&&\mbox{ }+\bigl(1-G(k+h-2i)\bigr)\bigl((k-i)G(k+h-2-2i)+iG(k+h+2-2i)\bigr) 
\biggr)
\label{psT}
\end{eqnarray}
\end{widetext}
Stationary value of overlap $x^*$ is obtained by means of fixed point 
equation
$x=F(x)$.
The existence of a phase transition depends uniquely on the nature of the
fixed point $x=1$. If it is attractive, two initial configurations 
differing by an infinitesimal fraction of nodes will become almost identical, 
whereas if it is repulsive, they will produce diverging trajectories.
A possible critical point is then determined by the equation 
\begin{eqnarray}
\frac{dF}{dx}|_{x=1}=1
\label{critical}
\end{eqnarray}
Let us recall briefly sufficient conditions for attractive fixed point $x^*$
, that is written as, Eq.(\ref{overlap}) satisfies 
(a)$dF/dx\geq0$ for all $0\leq x \leq 1,$ (b)$F(0)>0$ and (c)$dF/dx|_{x=1}>1$.
For $T=0$, $p_s$ is given by,
\begin{eqnarray}
\label{ps}
 p_s(k,h)  =\left\{
 \begin{array}{ll}
 \displaystyle{ k \choose (k+h)/2}/2^k,
  & \quad  \mbox{for $k+h$ is even}\\
 p_s(k-1,h),
  & \quad \mbox{for $k+h$ is odd}
 \end{array}\right.
\end{eqnarray} 
Moreover, $h$ is also set to zero. Item (a), (b) are manifest, and (c) is surprisingly satisfied for 
any $P(k)$ if $P(0)=0$. It is straight-forward to prove (c), that we show
$kp_s(k)-1\ge0, \mbox{for } \forall k\ge1$,
 using asymptotically $p_s(k) \sim k^{-1/2}$ for large $k$. 
In this case, fixed point $x=1$ is always repulsive and another fixed point $x^*<1$ is 
stable. That is, independent of the in-degree distribution $P(k)$,
there is no phase transition and a small perturbation/damage could propagate over 
the entire system. 
\begin{figure} 
\includegraphics{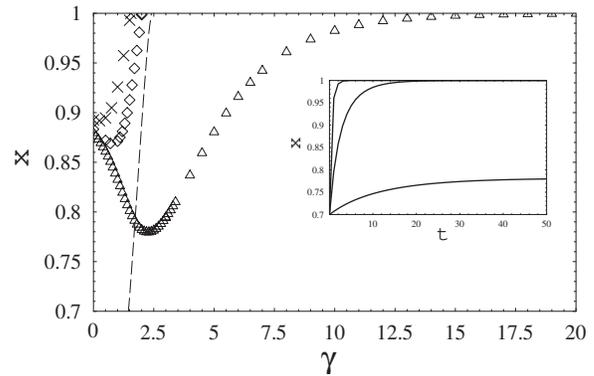}
\caption{
The normalized overlap function $x$ for $t\rightarrow \infty$
as a function of $\gamma$ with $h=0(\triangle)$,  $h=1(\diamond)$, 
$h=2(\times)$. The dashed line is that of RBN with $p=1/2$.
 Cutoff $\tilde{K}=100$ and temperature $T=0$. The critical $\gamma_c$ doesn't exist
(infinity) for $h=0$, 2.007 for $h=1$, 1.551 for $h=2$;
i.e., $\langle K_c\rangle=1(h=0)$, $\langle K_c\rangle=3.140(h=1)$, 
$\langle K_c\rangle=6.800(h=2).$ The inset shows time evolution of overlap 
function $x(t)$ with $x(0)=0.7$ and $\gamma=2.5$ for each threshold.}
\label{figure1} 
\end{figure} 
We now introduce the scale free in-degree distribution $k_i$ given by,
\begin{eqnarray}
P(k)&=&\left\{
 \begin{array}{ll}
 D,& \quad k=0\\
 (1-D)\eta(\gamma,\tilde{K})^{-1}k^{-\gamma},& \quad 1\le k \le \tilde{K}
 \end{array}\right.
\label{indegree} 
\end{eqnarray}
where $\eta(\gamma,\tilde{K})=\sum_1^{\tilde{K}}k^{-\gamma}$ with cutoff $\tilde{K}$, $0\le D \le1$ and $\gamma$ is usually called scale free exponent.
Substituting Eq.(\ref{indegree}) to Eq.(\ref{critical}), we obtain the critical condition for $D$.
\begin{eqnarray}
D_c=\frac{A(\gamma)-1}{A(\gamma)},\quad  A(\gamma)=\eta(\gamma,\tilde{K})^{-1}\sum_{k=1}^{\tilde{K}}p_s(k)k^{1-\gamma}
\label{Dc}
\end{eqnarray}
Since $1-p_s(k,0)\ge0$ for $\forall k\ge1$,
critical average connectivity $\langle K_c \rangle=\eta(\gamma-1,\tilde{K})/(A(\gamma)*\eta(\gamma,\tilde{K}))$
has a minimum value in the $\gamma\rightarrow \infty$, i.e. minimum critical average
$\langle K_c \rangle=1 $ is realized when $D_c=0$.

In Fig.$\ref{figure1}$, we plot the stationary overlap function $x^*$ for 
various threshold parameters, provided that $T=D=0$. 
The overlap function for the case $h=0$ is shown to asymptotically
approach $x^* \rightarrow 1$ with increasing $\gamma$, i.e.
average connectivity is approaching $\langle K \rangle=1$.
It is obviously different from RBN in which exists well known chaotic to 
ordered phase transition. 
For finite thresholds $h$, we observed equilibrium overlap is decreased as threshold incereases,
which suggests they eliminates the memory of initial condition and two different 
initial condition become identical.
For the comparison, The curve of equilibrium overlap for RBN is denoted.
One observes a pertuabation is rather suppressed in RTN with zero-threshold than in RBN for low $\gamma$ regime.
The opposite is observed for high $\gamma$ regime. The critical point where the 
strength of perturbation reverse is
given by $\gamma_s\approx 1.754$ in this model. This result supports the numerical
simulation observed in Fig.8 of \cite{RoBo}.
\begin{figure} 
\includegraphics{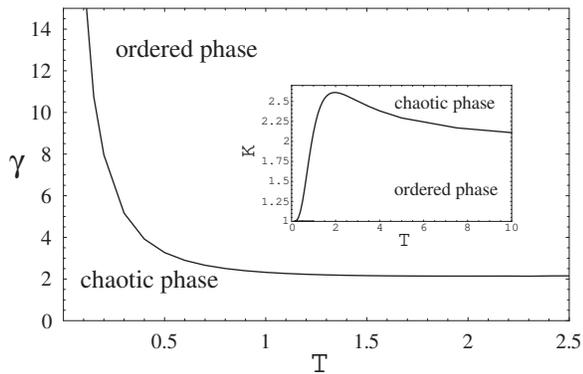}
\caption{Critical $\gamma_c$ curve as a function of $T$. In the chaotic 
($\gamma<\gamma_c(T)$) regime,
overlap function in the stationary state $x^*$ satisfies $x^* <1$. 
In the ordered ($\gamma>\gamma_c(T)$) regime, $x^* =1$. Threshold $h=0$ and 
cutoff $\tilde{K}=100$. The inset $\langle K\rangle-T$ diagram shows $\langle K_c \rangle$ 
curve has maximum value aroud $T=1.97$.}
\label{figure2} 
\end{figure} 

Fig.$\ref{figure2}$ shows the curve of critical scale free exponent 
$\gamma_c$ as a function of temperature for $h=D=0$, which distinguish ordered state from 
chaotic state. For finite temperature $T$, $dF/dx|_{x=1}$ in large $\gamma$ limit is given 
by,
\begin{eqnarray}
\frac{dF}{dx}| _{x=1_-}=&&\frac{1}{(1+\exp(-2/T))^2}\nonumber \\
+&&\frac{1}{(1+\exp(2/T))^2}
+\cal{O} (\mbox{1}/N)
\end{eqnarray}
The existence of infinitesimal $T$ which satisfies $dF/dx|_{x=1}<1$ results in critical temperature $T_c\simeq 0$.
The critical line would diverge in the zero temperature limit.
 An interesting feature of the $\gamma_c$ curve is that it has a minimum 
value $\gamma_c=2.1397$ , i.e. maximum $\langle K_c \rangle=$2.611, at 
$T=1.97\pm.003$.
The same phenomenon can be observed in other connectivity models, for 
example, Poisson distribution model where the in-degree distribution 
follows $P(k)=e^{-K_p} K_p^k/k!$.
Parameter $K_p$ is the average connectivity of the network.
It has the critical average connectivity $\langle K_c \rangle=1.8494$ at $T=0$ 
\cite{RoBo}.
In the low $T$ regime, the higher the temperature $T$, the higher the critical
average connectivity $\langle K_c\rangle$, and has a maximum value $\langle K_c \rangle=2.1994$ 
around $T=1.683\pm.001$ (subsequently goes to $\langle K_c \rangle=2$).
Both results show that moderate thermal noise can extend the regime of
ordered dynamics in RTN to higher average network connectivities, compared
to the zero temperature and the infinite temperature regime.
The infinite temprature $T\rightarrow \infty$ asymptotically leads to 
$p_s=1/2$, which is identical to RBN with $p=1/2$. Substituting it to Eq.(\ref{critical}),
the critical 
condition in RBN is given by,
\begin{eqnarray}
\frac{\eta(-1+\gamma,\tilde{K})}{2\eta(\gamma,\tilde{K})}=1
\label{Tinfty}
\end{eqnarray} 
Since average connectivity of the network is given by $\langle 
K\rangle=\eta(-1+\gamma,\tilde{K})/\eta(\gamma,\tilde{K})$,
one finds $\langle K_c \rangle=2$ from Eq.(\ref{Tinfty}).
Notice this value is independent of $\tilde{K}$ and $\gamma$.
Indeed, it is independent of any topological structures.
In the $\tilde{K}\rightarrow \infty$ limit, Eq.(\ref{Tinfty}) is 
represented by Riemann $\zeta$ function, which gives the critical value $\gamma_c \approx 
2.4788$.

In the following, we approach the system from another point of view,
by describing the single-site equation governing the time evolution of the node state. 
We show a simple example of agent's strategy game.
We first take the node state $\sigma_i(t)$ as agent's strategy at time $t$,
 i.e. $\sigma_i(t)=1$ as strategy $A$,  and $\sigma_i(t)=-1$ as strategy $B$.
One can define $\rho_k(t)$ as a density of agents which have connectivity $k$ and 
take strategy $B$ at time $t$.
We assume the interaction weight $c_{ij}$ takes only $1$ and $0$ and threshold $h$ to zero
to simplify the 
dynamics. This alteration does not change the dynamical overlap equation 
Eq.($\ref{psT}$).
This model can be also interpreted as a 'locally majority selection game'
which means each agent choose the strategy at time $t+1$ the same as 
the majority of their inputs at time $t$.
If the number of inputs strategies $A$ and $B$ is equal,
the agent choose either one with equal probability.
The dynamical mean field reaction rate equation for $\rho_k(t)$ is written as \cite{MaDi,SatVes}
\begin{eqnarray}
\partial_t \rho_k(t)=-\delta_k \rho_k(t)+\nu_k (1-\rho_k(t))
\label{rhoeq}
\end{eqnarray}
Agents with connectivity $k$ and strategy $A$ would change its strategy to $B$ with rate $\nu_k$, 
whereas agents with strategy $B$ would change to $A$ with rate $\delta_k$.
By using $\rho=\sum_k^{\tilde{K}}P(k)\rho_k$ which gives the 
average density of agents with strategy $B$, these rate parameters are written as 

\begin{eqnarray}
\delta_k
&=&
\sum_{i=0}^k {k \choose {i}}\rho^{i}(1-\rho)^{k-i}G(k-2i)\nonumber 
\\
\nonumber \\
\nu_k&=&1-\delta_k
\end{eqnarray}
Imposing the stationary condition on Eq.($\ref{rhoeq}$), we find stationary 
densities
$\rho_k^*=1-\delta_k^*$ since $\rho_k$ is on its turn function of 
$\delta_k$.
A self consistency equation that allows us to find $\rho^*$ is given by:
\begin{eqnarray}
\rho =1-\sum _{k=1}^{\tilde{K}} \sum_{i=0}^k {k \choose 
{i}}\rho^{i}(1-\rho)^{k-i}G(k-2i)P(k)
\label{selfrho}
\end{eqnarray}
From Eq.(\ref{selfrho}), An asymptotic $\gamma-T$ phase diagram similar to Fig.\ref{figure2} is obtained.
In $\gamma<\gamma_c(T)$ regime, the solution of Eq.($\ref{selfrho}$) has 3 fixed points 
$\rho=\rho_1(<1/2) ,1/2,\rho_2(>1/2)$. 
However, the fixed point at $\rho=1/2$ is an unstable solution because the 
derivative of right hand in Eq.($\ref{selfrho}$) has more than unity.
Whether the density $\rho$ goes to $\rho_1$ or $\rho_2$ in the stationary state 
depends on its initial condition.
In $\gamma>\gamma_c(T)$ regime, the equation has only one fixed and stable point at 
$\rho=1/2$, which
leads that any initial configurations asymptotically reach $\rho=1/2$ 
in the stationary state.
\begin{figure} 
\includegraphics{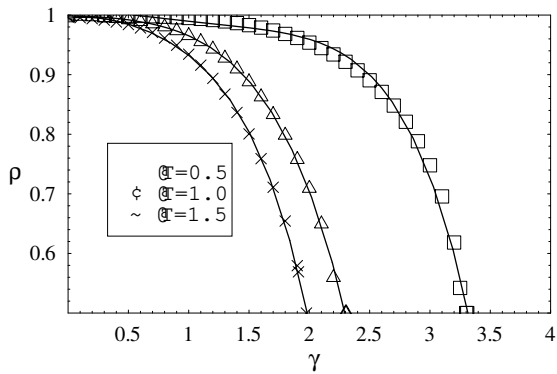}
\caption{The stationary density of nodes with strategy $A$ or $B$ as a 
function of $\gamma$ for various $T$. $\tilde{K}=100$.
The density $\rho$ is symmetry axis to $\rho=1/2$. The numerical 
simulation have been
run in networks of $N=10^5$ nodes and averaging over 50 different 
realizations.
The full and dash line are the corresponding analytical results.
Critical $\gamma_c$ which satisfies $\rho(\gamma)=1/2$ is 
$\gamma_c=3.309(T=0.5)$, $\gamma_c=2.307(T=1)$,
$\gamma_c=1.976(T=1.5)$,}
\label{figure3} 
\end{figure} 
In Fig.$\ref{figure3}$, we show the density $\rho$ in the 
upper plane ($\rho>1/2$) where the density $\rho_2$ reaches $\rho=1/2$ as $\gamma$ increases. 
On the other hand, in the $T\rightarrow 0$ limit, stationary density $\rho$ 
is given by,
\begin{eqnarray}
\lim_{T\rightarrow0} \rho_1 = 0,
\lim_{T\rightarrow0} \rho_2 = 1, \mbox{for finite $\gamma$}
\end{eqnarray}
The time interval required to reach the equilibrium state $t^*$ is 
estimated as $t^*\simeq e^{\gamma}$. Since it takes long time to reach stationary 
state for large $\gamma$, 
the network would be practically fixed to its initial state for large 
$\gamma$ limit.
In a small $\gamma$ area, the cutoff $\tilde{K}$ become crucial, which 
leads to
 $\lim_{T\rightarrow \infty}\rho_2(\gamma=0)=\tilde{K}/(\tilde{K}+1), 
\lim_{T\rightarrow 0}\rho_2(\gamma=0)=1$.
For finite $h$ case, $\rho=0$ is the only stable point 
for $\gamma>\gamma_c$, whereas there is another stable point for 
$\gamma<\gamma_c$.

In conclusion, we have studied the properties of Random Threshold Network 
model on scale free networks.
In $h=0, T=0, D=0$ case, we found there is no phase transition with respect to
average connectivity in any network topology.
, i.e. propagation of perturbation/damage independent of $\gamma$.
Regarding to $D$, it is quite interesting that minimum critical average connectivity is obtained when the rate of $k=0$ in-degree distribution is zero.
We also clarifies the relationship between RTN and RBN, 
our findings that moderate thermal noise extends the regime of ordered dynamics
to higher average connectivities, i.e. suppresses the propagation of perturbation/damage,
suggest a new aspect to e.g. the theory of neural networks.
Tuning thermal noise parameter in networks with asymmetric connections (as usually found in neural networks) where the basic units (neuron) follows threshold dynamics, thermal noise, 
i.e. non-zero error rate, may in fact improve information processing, 
given hierachical structure of the network topology.
 
\end{document}